# Addressable optical key distribution for unconditionally secured cryptography using phase-controlled quantum superposition

Byoung S. Ham

School of Electrical Engineering and Computer Science, Gwangju Institute of Science and Technology
123 Chumdangwagi-ro, Buk-gu, Gwangju 61005, S. Korea
(Submitted on November 21, 2018; bham@gist.ac.kr)

Based on the detection loophole-free photon key distribution (PKD) compatible with classical optical systems, an optical key distribution (OKD) protocol is presented for unconditionally secured cryptography in fiber-optic communications networks using addressable continuous phase basis, where each communication channel is composed of paired transmission lines. The unconditional security in OKD lies in quantum superposition between the paired lines of each channel. The continuous phase basis in OKD can be applied for one-time-pad optical cryptography in networks, whose network address capacity is dependent upon the robustness of OKD to channel noises.
DOI:

Due to the exponential growth of information traffic in fiber-optic backbone networks over the last thirty years, the theoretical limit of 100 Tbps in the current single-core fiber will be reached in a decade [1]. Then, multi-core fibers should replace the current single-core fibers to keep ever-expanding IT technologies on track [1]. This multicore fiber satisfies phase stability in March-Zehnder interferometer (MZI) due to its structural robustness [2]. Current information security in the Internet is mostly based on public key cryptography relying on computational complexity, resulting in the inherent conditional security [3]. Thus, the Internet security is vulnerable to computational power. For the unconditional security from channel eavesdropping, quantum cryptography [4] has been studied for quantum key distribution (QKD) over the last thirty years since the first advent of QKD protocol of BB84 [5]. In practice, however, QKD security is also fragile to eavesdropping due to imperfect detectors and channel noises, resulting in a detection loophole [6]. This practical issue of detection loophole puts quantum cryptography in danger even for decoy states [7] for single photons as well as Bell stats [8] for entangled photon pairs. For transmission distance, QKD is strongly limited by the no cloning theorem [9,10]. Quantum networking requires multipartite entangled photon pairs which is much harder to be implemented than the point-to-point QKD [11]. Based on these practical limitations, quantum cryptography seems to have a long way to go for commercial applications such as online banking, IoT, and cryptocurrency [12]. Most of all, no QKD protocol is compatible with current fiber-optic communications simply due to the no cloning theorem as well as nonclassical light usage.

To solve the limitations in both classical and quantum cryptographies, a photon key distribution (PKD) protocol has been proposed recently for the detection loophole-free unconditional security using bright coherent lights in paired transmission lines composing a MZI for a single channel [13]. The detection loophole-free is from the transmission determinacy in MZI coherence optics. The unconditional security of PKD is based on quantum superposition between the paired transmission lines, resulting in complete randomness in eavesdropping. According to the information theory, the complete randomness satisfies perfect security [14]. Although Eve's channel measurements are fully allowed in PKD, the measured information cannot be distinguished for the discrete keys of '0' and '1' due to the superposition-caused indistinguishability, where absolute phase information cannot be measured unless the input phase is known. The input phase information in PKD belongs to the transponder, which is beyond the channel security issue. Owing to coherence optics of MZI [15,16], PKD is naturally compatible with current optical systems such as optical switches, routers and even amplifiers. For the optical amplifiers, the phase between the input and output can be locked technically, so that the physics of MZI is kept for the PKD protocol [17]. Owing to the use of bright coherent light pulses as in the fiber-optic communications, the key distribution process in PKD is fully deterministic as in the classical information processing [13]. Thus, both bit rate and bit error rate of PKD should be compatible with those in current fiber-optic communications. The multicore fiber potentially used for the MZI channel is robust to environmental noises such as temperatures and vibrations owing to the proximity between cores within a few microns, where a relative path length change plays a key role in PKD.

In this paper, an optical key distribution (OKD) protocol is presented for the detection loophole-free, unconditional security compatible with current fiber-optic communications networks, where OKD is an addressable PKD by using continuous phase basis (CPB). Compared with a fixed point-to-point transmission scheme in PKD, the addressability in the present OKD is essential for networking. The availability of CPB in OKD is owing to the paired phase matching condition in the MZI channel controlled by two sets of phase controllers as shown in Fig. 1. The physics of the unconditional security in OKD is in the MZI path superposition like in PKD, completely different from all QKD protocols using a single channel governed by no-cloning theorem of dual bases. The continuity of the phase basis in the present OKD can be directly applied to addressable key distribution process for dense wavelength division multiplexing (DWDM) in the future multicore fiber-optic networks [18].

Figure 1 shows a schematic of the proposed OKD based on double phase-controlled MZI between two remote parties, Alice and Bob. Each party has each phase shifter to encode one's optical keys with a phase $\varphi_1$ for Φ1 (Bob) and $\psi_1$ for Ψ1 (Alice), respectively. The MZI scheme looks similar to the phase encoded BB84 protocol [19], but completely different in the use of two transmission lines for a single channel and deterministic key distribution technique for a round trip. Most of all, the physics of unconditional security in OKD does not rely on dual bases resulting in the no cloning theorem [4-10]. The added phase controller Φ2 (Ψ2) is to control the encoder Φ1 (Ψ1) for its addressing in networks. In PKD without Φ2 and Ψ2 [13], the MZI satisfies unitary transformation if $\varphi_1 = \psi_1$ is satisfied for a round trip, where $\varphi_1$ and $\psi_1$ have the same set of two discrete bases: $\varphi_1, \psi_1 \in \{0, \pi\}$. Here, we briefly seek a new condition for the unitary transformation in the OKD scheme of Fig. 1. Here, the phase basis '0' ('$\pi$') represents a key '0' ('1').

The matrix representation, [BH], for the phase-controlled round-trip carrier-light pulse in Fig. 1 is as follow (see the Supplemental Material):

$$[BH] = \frac{1}{2}\begin{bmatrix} -\{e^{i(\psi_2+\varphi_1)} + e^{i(\psi_1+\varphi_2)}\} & i\{e^{i(\psi_2+\varphi_1)} - e^{i(\psi_1+\varphi_2)}\} \\ i\{e^{i(\psi_1+\varphi_2)} - e^{i(\psi_2+\varphi_1)}\} & -\{e^{i(\psi_2+\varphi_1)} + e^{i(\psi_1+\varphi_2)}\} \end{bmatrix}, \qquad (1)$$

where, the output light must satisfy the identity and inversion relation for key distribution: $\begin{bmatrix} E_9 \\ E_{10} \end{bmatrix} = [BH]\begin{bmatrix} E_1 \\ 0 \end{bmatrix}$. For this, Bob randomly prepares potential keys and sent them to Alice. Alice randomly chooses her phase basis for $\psi_1$ to encode the return light and sends it back to Bob. If the return light $E_9$ ($E_{10}$) hits the detector $B_3$ ($B_4$), it means the identity (inversion) satisfying a raw key (see the Supplementary Material). Unlike QKD, this key distribution is fully deterministic without sifting process due to no use of two nonorthogonal bases (dual bases) [13]. As a result, the following condition is obtained from equation (1):

$$\psi_1 + \varphi_2 = \psi_2 + \varphi_1, \qquad (2)$$

where the identity/inversion relation is used for raw keys (discussed in Table 1). For deterministic key distribution via transmission directionality in MZI physics, phase matching ($\psi_1 = \varphi_1$) should be satisfied as discussed in PKD (see also Fig. 2) [13]. Thus, equation (2) results in $\psi_2 = \varphi_2$ in the OKD scheme of Fig. 1. However, equation (2) lacks the relation between $\varphi_1$ and $\varphi_2$, so does $\psi_1$ and $\psi_2$. Keeping this in mind, we investigate CPB property of OKD for the unconditionally secured networking in a classical regime.

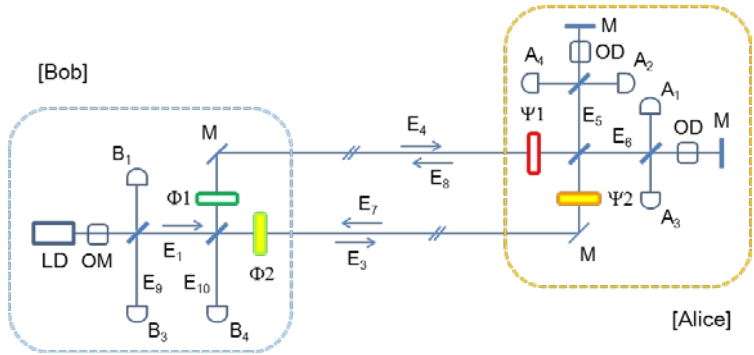

Fig. 1. A schematic of cv PKD. LD, Laser diode; OM, Optical modulator; Φ1/Φ2/Ψ1/Ψ2, Phase shifter; $A_1 \sim A_4/B_1/B_3/B_4$, Photodetector, OD, Optical delay, M, Mirror, $E_i$, Light $_i$. This distance between Bob and Alice is less than or equal to 100 km defined by EDFA in fiber optics communications if there is no coherent amplifications, otherwise unlimited (beyond the current scope).

The (bright) input light pulse $E_1$ in Fig. 1 is launched from a coherent laser (LD) through an optical modulator (OM) by Bob for a random phase basis $\varphi_1 \in \{0, \pi\}$ controlled by the phase shifter $\Phi_1$. Suppose that Fig. 1 represents one channel out of NxN network channels. The added control phase shifter $\Phi_2$ with control phase $\varphi_2$ ($0 \leq \varphi_2 \leq \pi$) is used to address the $\varphi_1$−controlled $E_1$, such that only a particular receiver should work for the MZI determinacy in OKD by the corresponding control phase $\psi_2$ at the phase shifter $\Psi_2$ (discussed below). Here, the MZI determinacy represents the phase-dependent transmission directionality: If $\varphi_1 = 0$ ($\varphi_1 = \pi$) assuming no relative phase shift between two lines of MZI, the detector $A_1$ ($A_2$) always clicks for $E_6$ ($E_5$) for $\varphi_2 = \psi_2 = 0$: The phase $\psi_1$ (as well as $\psi_2$) is invisible to the forward transmission of $E_3$ and $E_4$ [12]. Such fast optical switching is technically available via detector $A_i$-triggered delay operation. The $\psi_1$−controlled return light $E_8$ by Alice is also governed by the same MZI transmission directionality, resulting in identity/inversion relation (discussed in Figs. 2 and 3). To the return lights ($E_7$ and $E_8$), both phases $\varphi_1$ and $\varphi_2$ are invisible, too.

Figure 2 shows numerical calculations of the MZI determinacy (transmission directionality) in Fig. 1 for the output lights $E_5$ and $E_6$ at Alice's side. The related matrix representation $[MZ]_{\varphi_1,\varphi_2}$ of the directionality for $E_5$ and $E_6$ is as follow:

$$[MZ]_{\varphi_1,\varphi_2} = \frac{1}{2}\begin{bmatrix} e^{i\varphi_2} - e^{i\varphi_1} & i(e^{i\varphi_2} + e^{i\varphi_1}) \\ i(e^{i\varphi_2} + e^{i\varphi_1}) & -(e^{i\varphi_2} - e^{i\varphi_1}) \end{bmatrix}, \quad (3)$$

where $\begin{bmatrix} E_5 \\ E_6 \end{bmatrix} = [MZ]_{\varphi_1,\varphi_2} \begin{bmatrix} E_1 \\ 0 \end{bmatrix}$. In Fig. 1, the added phase $\varphi_2$ causes a phase change in the lower transmission line for the forward light $E_3$. To compensate it, $\varphi_1$ has to be adjusted accordingly for $E_4$ in the upper transmission line. Thus, the modified phase at $\Phi 1$ must be $\varphi_1 \to \varphi_1 + \varphi_2$. With this modified phase, equation (3) can be easily proved for the MZI directionality for an arbitrary value of $\varphi_2$.

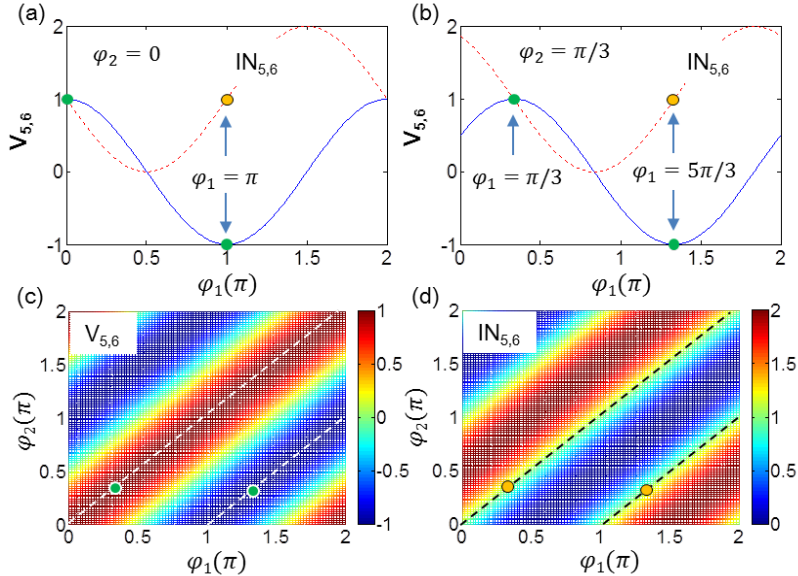

Fig. 2. Numerical calculations for the transmission directionality in MZI. Visibility $V_{5,6}$ (solid) and Interference $IN_{5,6}$ (dotted) for (a) $\varphi_2 = 0$ and (b) $\varphi_2 = \pi/3$. (c) $V_{5,6}$ and (d) $IN_{5,6}$. $V_{i,j} = \left(\frac{I_j - I_i}{I_i + I_j}\right)$, where $I_i$ is intensity of $E_i$. $IN_{5,6} = (E_5 + E_6) \cdot (E_5 + E_6)^*$.

For the numerical proofs of the $\varphi_2$−dependent MZI directionality mentioned above, two basis values of $\varphi_1 \in \{0, \pi\}$ are used to test both the visibility $V_{5,6}$ and interference $IN_{5,6}$. Here, the interference $IN_{5,6}$ should be the same as $IN_{3,4}$, representing Eve's measurement in the channel using the same measurement setup of Alice's (see the Supplementary Material). Figure 2a shows numerical results for $\varphi_2 = 0$ as a reference, while Fig. 2b is for an arbitrary value of $\varphi_2 = \pi/3$. As a typical example of the PKD protocol, two phase bases of $\varphi_1 \in \{0, \pi\}$ show the maximum visibilities of $V_{5,6} = \pm 1$ in Fig. 2a (see the green dots in the solid curve), while the interference, $IN_{5,6} = 1$, shows indistinguishability with the same value (see the green and orange dots in the dotted curve) [13]. As discussed in ref. 13, this $IN_{5,6}$ is the same as $IN_{3,4}$, showing the physical origin of the measurement immunity in the MZI paths: unconditional security (discussed later). The compensated $\varphi_1$ according to the control phase $\varphi_2$ is shown in Fig. 2b: $\varphi_1 \to \varphi_1 + \pi/3(\varphi_2)$. For the maximum visibility $V_{5,6} = \pm 1$, the phase matching condition must be satisfied: $\varphi_1 = \varphi_1 + \varphi_2$. This linear phase matching relation between $\varphi_1$ and $\varphi_2$ reveals the infinite number of phase choices on $\varphi_2$, resulting in the continuous CPB characteristics of OKD as shown in Fig. 2c. In other words, the control phase $\varphi_2$ can be used for addressing the $\varphi_1-$ provided keys to a specific destination at Alice's side. The corresponding interference $IN_{5,6}$ has always the same value if $\varphi_1 = \varphi_1 + \varphi_2$ is satisfied as shown in Fig. 2d. Thus, Fig. 2 proves the $\varphi_2$−dependent MZI determinacy in the OKD scheme of Fig. 1 as well as indistinguishability in eavesdropping (discussed later). The addressable condition of OKD is $\varphi_1 = \varphi_1 + \varphi_2$.

Because $\varphi_1 = \psi_1$ must be satisfied for the one-way deterministic key transmission in OKD, $\psi_2$ at Alice's side should be equal to $\varphi_2$ in the same way to compensate the shifted $\psi_1$ in Fig. 2b (see also equation (2) and the Supplementary Material). Figure 3 shows the numerical calculations with addressable CPB variables for $\varphi_2 = \psi_2$ in the present OKD. To satisfy the identity matrix at Bob's side for equation (1), the visibility of $V_B = -1$ for both bases ($\varphi_1 = \psi_1 = \{0, \pi\}$) is numerically proved in Fig. 3a for an arbitrary value of $\varphi_2 = \psi_2 = 2\pi/5$: $V_B = V_{9,10}$. Thus, the phase basis set of $\varphi_1 \in \{0, \pi\}$ becomes continuous because of $0 \leq \varphi_2 \leq \pi$, where each $\varphi_2$ represents a specific address for networking: $\varphi_2 = \psi_2$. Practically the possible number of CPB for the MZI channel is determined by the detector's sensitivity. This addressable condition ($\varphi_2 = \psi_2$) seems to be obvious if no path-length (phase) change occurs in the MZI channel. In general, the invariant MZI path length can be easily achieved even for independent two single-core fibers by using, e.g. a laser locking technique due to slow noise generation process [20-22]. Thus, the MZI stability is just a technical issue.

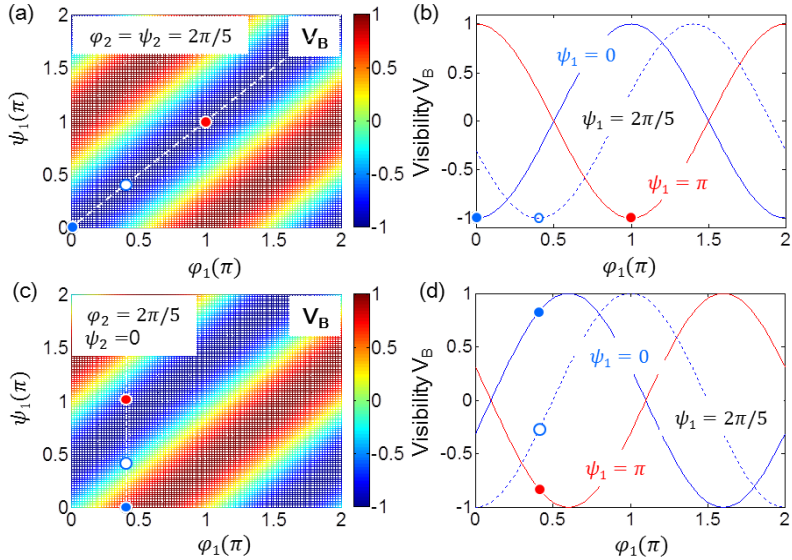

Fig. 3. Numerical calculations of visibility $V_B$ for OKD. Visibility $V_B$ (a) and (b) for $\psi_2 = \varphi_2 = 2\pi/5$, and (c) and (d) for $\psi_2 = 0$ and $\varphi_2 = 2\pi/5$. Calculations are based on equation (1). $V_B = \left(\frac{I_{10}-I_9}{I_{10}+I_9}\right)$: $I_i$ is the intensity of $E_i$.

Figure 3b shows the $\psi_1$−independent identity relation ($\psi_1 = 0; \frac{2\pi}{5}; \pi$) for the phase matching in Fig. 3a. Although the dashed curve is the only $\varphi_2$−corresponding $\psi_1$ for addressable key distribution, all $\psi_1$ values are allowed if $\psi_1 = \varphi_1$ is satisfied (see the blue and red dots). However, the directionality $V_{5,6}$ is broken if $\varphi_1 \neq \varphi_1 + \varphi_2$ (see Fig. 2b): $V_{5,6} \neq -1$. Thus, only the dotted curve with $\psi_1 = \frac{2\pi}{5}(=\varphi_2)$ satisfies the determinacy condition with $V_{5,6} = -1$. This is because $\varphi_1$ is shifted by the $\varphi_2$ value, and the shifted $\varphi_1$ bases affect $\psi_1$ to keep $V_{5,6} = \pm 1$ (directionality/determinacy). Then, how does Alice know about her $\psi_1$ adjustment? This is the power of the addressable networking of CPB in OKD when a $\varphi_2$ value is shared with a particular $\psi_2$ for a specific address. If $\varphi_2 \neq \psi_2$ for a wrong address set, the identity relation ($V_B = -1$) is broken as shown in Figs. 3c and 3d. For the addressable OKD, Bob provides $\varphi_2-$ dependent $\varphi_1$ for the maximum visibility $V_{5,6}$ measured by Alice as shown in Fig. 2b: $\varphi_1 = \varphi_1 + \varphi_2$. If Alice does not know the adjusted $\varphi_1$ by Bob, then she may randomly choose her basis $\psi_1 \in \{0, \pi\}$ without adjustment. As a result, the visibility $V_B$ (=$V_{9,10}$) for the return light does not satisfy the identity relation as shown in Fig. 3d: $V_B \neq -1$. Here, $V_B \neq -1$ means that the detector $B_4$ is also clicked on for $E_4$ indicating an error. Unless Alice knows Bob's $\varphi_2$ value, therefore, the key distribution process is failed. This is the addressable CPB property, where $\varphi_2$ represents the address at Bob's side in OKD networking. The $\varphi_2$-corresponding address at Alice's side is $\psi_2$.

Figure 4 shows numerical calculations of channel measurements in the MZI paths for the proof of unconditional security in OKD. The matrix representation $[MZ]_{\psi,\varphi}$ for both $E_7$ and $E_8$ in the MZI paths of Fig. 1 is denoted by:

$$[MZ]_{\psi,\varphi} = \frac{1}{\sqrt{2}} \begin{bmatrix} -e^{i(\psi_2+\varphi_1)} & ie^{i(\psi_2+\varphi_1)} \\ ie^{i(\psi_1+\varphi_2)} & -e^{i(\psi_1+\varphi_2)} \end{bmatrix}, \quad (4)$$

where $\begin{bmatrix} E_7 \\ E_8 \end{bmatrix} = [MZ]_{\psi,\varphi} \begin{bmatrix} E_1 \\ 0 \end{bmatrix}$ (see the Supplementary Material). Figures 4a and 4b show both interference $IN_{7,8}$ and visibility $V_{7,8}$ allowing Eve's perfect measurements without altering the outputs at Bob's detectors. The channel intrusion by Eve without altering the output fringe is challenging but not impossible. Even in this case of perfect measurement, however, knowing the absolute phase information of the light carrier is definitely impossible, unless $E_1$ information is given to Eve in advance. Here, the purpose of eavesdropping is to know the phase information of the light for keys. Thus, the path superposition in the MZI for coherence optics corresponds to the no-cloning theorem in QKD based on nonorthogonal basis sets. According to equation (2), Alice's phase adjustment on $\psi_1$ must be the same as the Bob's adjustment on $\varphi_1$ due to the condition of $\psi_2 = \varphi_2$ for directionality. In this balanced case, the Eve's channel attacking becomes failed due to indistinguishable results for two keys as demonstrated in Figs. 4a and 4b. This indistinguishability roots in equation (4), where the satisfaction with equation (2) requires the same value in the exponents of the matrix elements. Even for the address mismatch ($\psi_2 \neq \varphi_2$), the measurement randomness in the channel is still effective as shown in Figs. 4c and 4d. In this unbalanced case, the value of $IN_{7,8}$ is just shifted by $\varphi_2 (\neq \psi_2)$ compared to that in Fig. 4(a) according to equation (4). Thus, the randomness or indistinguishability in MZI path measurements by Eve is sustained for all $\varphi_2$-dependent network channels.

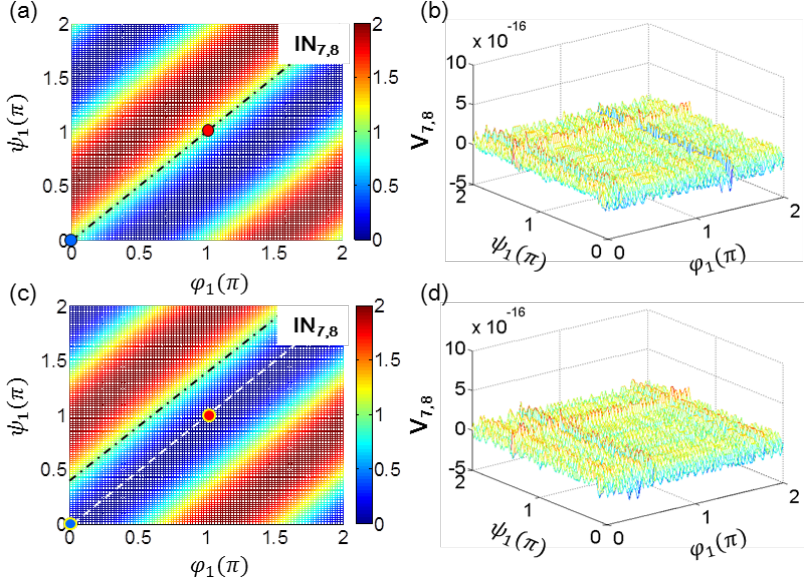

Fig. 4. Numerical calculations of interference $IN_{7,8}$ and visibility $V_{7,8}$ for Fig. 1. (a) $IN_{7,8}$ and (b) $V_{7,8}$ for $\psi_2 = \varphi_2 = 2\pi/5$. (c) $IN_{7,8}$ and (d) $V_{7,8}$ for $\psi_2 = 0$ and $\varphi_2 = 2\pi/5$. The keys are denoted by dots: $\varphi_1 = \psi_1 = 0$ (blue); $\varphi_1 = \psi_1 = \pi$ (red). $IN_{7,8} = (E_7 + E_8)(E_7 + E_8)^*$; $V_{7,8} = \left(\frac{I_8 - I_7}{I_7 + I_8}\right)$.

Regarding the eavesdropping discussed in Fig. 4, however, Eve can set up exactly the same measurement tools as Alice and Bob have for both outbound and inbound lights, respectively. Then, Eve simply reads out the visibility as Alice and Bob do. Because the initial MZI condition is not known, Eve's best chance to guess is 50% of Alice's even with careful phase adjustment for the maximum values: $|V_{3,4}| = |V_{7,8}| = 1$. Thus, both one-way determinacy and the round-trip identity/inversion relations in the phase controlled MZI cannot be satisfied to eavesdropping, resulting in unconditional security even with perfect channel intrusion. This is the fundamental difference of OKD from others such as QKD. The unconditional security in OKD is based on the path superposition-caused indistinguishability, where the unconditional security in QKD roots in the Heisenberg's uncertainty principle applied to dual bases (two nonorthogonal bases) resulting in the no cloning theorem. However, Eve may try an offline attack with the recorded data based on coherence optics by simply flipping each measured bit value if the decoded data have no meaning. This type of post-measurement attack is still of no use if the initialization process between Bob and Alice is randomly repeated to reset the MZI such as done with $\varphi_2$ in Fig. 3. If the MZI is under active phase locking [20-22], Eve's post-measurement attack is completely random for each bit due to random phase fluctuations between the original MZI channel and the Eve's tapping setup for eavesdropping.

Table 1. Sequence for OKD in Fig. 1. The phase $\varphi_1$ is denoted without addition of $\varphi_2$. So does $\psi_1$.

| Party | Order / Sequence | | 1 | 2 | 3 | 4 | 5 | 6 | 7 | 8 | 9 | 10 | set |
|---|---|---|---|---|---|---|---|---|---|---|---|---|---|
| Bob | 1 | $\varphi_1$* | 0 | 0 | $\pi$ | 0 | $\pi$ | $\pi$ | 0 | $\pi$ | 0 | $\pi$ | |
| | 2 | Prepared key: $x(\varphi_1)$ | 0 | 0 | 1 | 0 | 1 | 1 | 0 | 1 | 0 | 1 | $\{x\}$ |
| | 8 | $V_B$ | 1 | −1 | 0.9 | 1 | −1 | −1 | −1 | 1 | 1 | 1 | |
| | 9 | Raw key | **0** | **1** | **X** | **0** | **1** | **1** | **1** | **0** | **0** | **0** | $\{m_B\}$ |
| | 10 | Final key | **0** | **1** | **X** | **0** | **1** | **X** | **1** | **0** | **0** | **0** | $\{m\}$ |
| Alice | 3 | $V_A$ | 1 | 1 | −1 | 1 | −1 | −0.8 | 1 | −1 | 1 | −1 | |
| | 4 | Copy x: y | 0 | 0 | 1 | 0 | 1 | −0.8 | 0 | 1 | 0 | 1 | $\{y\}$ |
| | 5 | $\psi_1$ | $\pi$ | 0 | 0 | $\pi$ | $\pi$ | $\pi$ | 0 | 0 | $\pi$ | 0 | |
| | 6 | $z(\psi_1)$ | 1 | 0 | 0 | 1 | 1 | 1 | 0 | 0 | 1 | 0 | $\{z\}$ |
| | 7 | Raw key | **0** | **1** | **0** | **0** | **1** | **X** | **1** | **0** | **0** | **0** | $\{m_A\}$ |
| | 10 | Final key | **0** | **1** | **X** | **0** | **1** | **X** | **1** | **0** | **0** | **0** | $\{m\}$ |

Table 1 shows the key distribution sequence in the present OKD. Bob simply prepares a random bit sequence for Alice by using his phase basis $\varphi_1$. Alice also randomly encodes the Bob-prepared bits for raw keys using her phase bases $\psi_1$. Owing to the determinacy and identity relations, the raw key is deterministically obtained by simply reading out the visibilities and comparing them with their phase selections without sifting if there is no error. Unlike PKD [13], both identity and inversion are used for keys to double the key distribution rate (see the Supplementary Material). For error corrections, both parties publically announce their error bits (red numbers) and remove them from the raw key. As a result, the same length final key (m) is shared. Here, the mark X means a discarded bit. To evaluate an error rate, Alice and Bob randomly pick some samples out of the final key sequence and compare them: privacy amplification. The randomly chosen bits for the privacy amplification are removed from the final key. Thus, the bit rate in OKD is strongly dependent upon the bit error rate as usual. The followings are the key distribution sequence of OKD shown in Table. 1:

1. Bob randomly selects his phase basis $\varphi_1 \in \{0, \pi\}$ to prepare a key and sends it to Alice.
2. Bob converts the chosen basis $\varphi_1$ into a key for his key record x: x∈ {0,1}, if $\varphi = 0$, x=0; if $\varphi = \pi$, x=1.
3. Alice measures her visibility $V_A$ and keeps the record.
4. Alice copies the Bob's key for her record y: if $V_A$=1, y=0; if $V_A$=−1, y=1; if $V_A \neq \pm 1$, y=$V_A$ (error).
5. Alice randomly selects her basis $\psi_1 \in \{0, \pi\}$, encode the return light, and sends it back to Bob.
6. Alice converts the chosen basis $\psi_1$ into a key record z: z ∈ {0,1}; if $\psi_1 = 0$, z=0; if $\psi_1 = \pi$, z=1.

7. Alice compares y and z for the raw key $m_A$: $m_A = (y + z) \oplus 1$ at modulus 2. If $m_A \neq \{0,1\}$, $m_A = X$ (error).
8. Bob measure his visibility $V_B$.
9. Bob set the raw key $m_B$: if $V_B = 1$, $m_B = 0$; $V_B = -1$, $m_B = 1$. if $V_B \neq \pm 1$, $m_B = X$ (error).
10. Alice and Bob publically announce their error bits and remove them from their raw keys to set the shared final key m.

Lastly, the present OKD can be applied to 1xN and NxN network configurations (see the Supplementary Material). As discussed in Figs. 1~4, each $\varphi_2$−dependent phase shift in the $\varphi_1$ basis can be individually allocated to each wavelength in the DWDM system of fiber-optic communications networks [18], in which the single-core fibers will be replaced by multi-core fibers in the near future. In such a multi-core fiber, the MZI path length is potentially distance-unlimited due to the environmental noise robustness, where temperature, vibration and turbulence are generally the major contributors to the dynamic path-length shift. The laser fluctuation in both intensity and phase is also independent of the measurement sensitivity due to the benefit of the MZI coherence optics, where ultra-stable detection sensitivity can be provided by even a low-quality laser system. For the case of, e.g., a 10 Gpbs key rate at each OKD channel, the coherence length $l_C$ is calculated for $l_C = \frac{c}{\Delta} = 3\ cm$ in free space. Such a coherence length is long enough even to a bulky system for a carefully controlled visibility measurement at transponders. The wavelength converter, optical MUX/DEMUX, and an amplifier such as EDFA are coherent devices, so a phase difference between input and output can be locked. This fixed phase shift can also be technically adjusted to have the same visibility between Bob and Alice in a network preparation stage for address allocation. For a wavelength sharing network configuration, STAR or FTTH fiber optic network is also possible. Thus far, we have analyzed and discussed OKD in a phase-controlled MZI system for unconditionally secured and deterministic optical key distribution between remote two parties, where the networking property of the control phase $\varphi_2$ can be used for addressing the network channels by allocating a specific phase ($\varphi_2$) value to a particular remote party ($\lambda$) as its network address.

In summary, an addressable OKD protocol was proposed, analyzed, and discussed for unconditionally secured and deterministic optical key distribution in network environments using a phase-controlled Mach Zehnder interferometer, in which the key carrier is bright coherent light and the channel measurement randomness is due to the quantum superposition between two transmission lines of the MZI channel. Compared with current QKD protocols such as BB84 based on single photons in a single transmission line, the proposed OKD uses bright coherent light on paired transmission lines composing MZI, resulting in detection loophole-free, ultrafast, and distance unlimited optical key distribution. Unlike dual bases-based no-cloning theorem in QKD, the physics of unconditional security in OKD lies in the quantum superposition between paired transmission lines of the MZI channel, resulting in randomness in channel eavesdropping. Owing to the control phase $\varphi_2$, OKD supports not only a point-to-point transmission scheme but also network configurations. Thus, the OKD can be applied to current optical systems in fiber-optic communications networks. Especially in the future multi-core fiber networks, the infrastructure of the MZI is fulfilled for the present OKD, where the multi-core fiber is immune to the environmental noises owing to the proximity among cores on a few micron scale, resulting in a potentially unlimited transmission distance. Even in the present single-core fiber systems, a real-time phase control for the MZI scheme was already shown successfully [22]. Thus, the present OKD may be applied for the long lasting goal of one-time-pad cryptography in a classical regime for ultrahigh speed fiber-optic communications networks [14].

The author acknowledges that the present work was supported by the ICT R&D program of MSIT/IITP (1711042435: Reliable crypto-system standards and core technology development for secure quantum key distribution network).

# Supplemental Material

for "Addressable optical key distribution for unconditionally secured cryptography using phase-controlled quantum superposition,"
by Byoung S. Ham

Gwangju Institute of Science and Technology, S. Korea
Email:bham@gist.ac.kr

Using equation (1), the following identity and inversion matrices are obtained for the round trip of light in Fig. 1:

$$\begin{bmatrix} E_9 \\ E_{10} \end{bmatrix} = [BH] \begin{bmatrix} E_1 \\ 0 \end{bmatrix} = \frac{1}{2} \begin{bmatrix} -\{e^{i(\psi_2+\varphi_1)} + e^{i(\psi_1+\varphi_2)}\} & i\{e^{i(\psi_2+\varphi_1)} - e^{i(\psi_1+\varphi_2)}\} \\ i\{e^{i(\psi_1+\varphi_2)} - e^{i(\psi_2+\varphi_1)}\} & -\{e^{i(\psi_2+\varphi_1)} + e^{i(\psi_1+\varphi_2)}\} \end{bmatrix} \begin{bmatrix} E_1 \\ 0 \end{bmatrix}, \quad (S1)$$

where $[BH]=\frac{1}{4}[BS][\psi_{1,2}][BS][BS][\varphi_{1,2}][BS]$. The matrices $[BS]$, $[\psi_{1,2}]$, and $[\varphi_{1,2}]$ are for the beam splitter, control phase shifter $\Psi_i$, and base phase shifter $\Phi_i$ in the MZI:

$$[BS] = \frac{1}{\sqrt{2}} \begin{bmatrix} 1 & i \\ i & 1 \end{bmatrix}, \quad (S2)$$

$$[\psi_{1,2}] = \begin{bmatrix} e^{i\psi_2} & 0 \\ 0 & e^{i\psi_1} \end{bmatrix}, \quad (S3)$$

$$[\varphi_{1,2}] = \begin{bmatrix} e^{i\varphi_2} & 0 \\ 0 & e^{i\varphi_1} \end{bmatrix}. \quad (S4)$$

(i) Identity relation:

From equation (S1), $E_9 = cE_1$ and $E_{10} = 0$ are satisfied, where c is a global phase factor. Thus, the exponent of each matrix element in [BH] must be:

$$(\psi_2 + \varphi_1) = (\psi_1 + \varphi_2). \quad (S5)$$

Because $\varphi_1, \psi_1 \in \{0, \pi\}$ and $\varphi_1 = \psi_1$ for the identity relation in PKD (see ref. 18), the following relation is obtained for the control phase:

$$\psi_2 = \varphi_2. \quad (S6)$$

Unlike phase bases $\psi_1$ and $\varphi_1$ having the same discrete value either 0 or π, equation (S6) is not restricted. Thus, the values of the control phases of $\psi_2$ and $\varphi_2$ are continuous between 0 and π. Regardless of the control phase value working for the continuous variables, the initially chosen base phase is invariant to the result.

(ii) Inversion relation:

From equation (S1), $E_9 = 0$ and $E_{10} = cE_1$ are satisfied. Thus, the exponent of each matrix element in [BH] must be:

$$(\psi_2 + \varphi_1) = (\psi_1 + \varphi_2) \pm \pi. \quad (S7)$$

Because $\varphi_1, \psi_1 \in \{0, \pi\}$ and $\varphi_1 = \psi_1 \pm \pi$ for the inversion in PKD (see ref. 18), the following relation is achieved:

$$\psi_2 = \varphi_2. \tag{S8}$$

Therefore, the control phase relation in equations (S6) and (S8) is universal either it is for identity or inversion. For the key distribution process in Fig. 1, Alice randomly selects her phase basis $\psi_1$ to be either identical or opposite to the Bob's choice. If Alice's basis-phase choice is identical to the Bob's, it results in the identity relation of equation (S5) regardless of the control phase if $\psi_2 = \varphi_2$, otherwise results in the inversion relation of equation (S7). In any case, the equality relation in equations (S6) and (S8) for the control phase provides the same result as PKD. In summary, any value of the control phase satisfies OKD for the photon key distribution using coherent light if $\psi_2 = \varphi_2$, and the infinite number of control phase values can be applied for infinite number of addresses in OKD. This is the theoretical background of the present continuous variables in OKD.

The matrix representation of $\psi_1$-controlled return light by Alice (see equation (4)) in the MZI channel of Fig. 1 is denoted by:

$$\begin{bmatrix} E_7 \\ E_8 \end{bmatrix} = \frac{1}{2\sqrt{2}} [\psi_{1,2}][BS][BS][\varphi_{1,2}][BS]\begin{bmatrix} E_1 \\ 0 \end{bmatrix}$$

$$= \frac{1}{2\sqrt{2}} \begin{bmatrix} e^{i\psi_2} & 0 \\ 0 & e^{i\psi_1} \end{bmatrix}\begin{bmatrix} 1 & i \\ i & 1 \end{bmatrix}\begin{bmatrix} 1 & i \\ i & 1 \end{bmatrix}\begin{bmatrix} e^{i\varphi_2} & 0 \\ 0 & e^{i\varphi_1} \end{bmatrix}\begin{bmatrix} 1 & i \\ i & 1 \end{bmatrix}\begin{bmatrix} E_1 \\ 0 \end{bmatrix}$$

$$= \frac{1}{\sqrt{2}} \begin{bmatrix} -e^{i(\psi_2+\varphi_1)} & ie^{i(\psi_2+\varphi_1)} \\ ie^{i(\psi_1+\varphi_2)} & -e^{i(\psi_1+\varphi_2)} \end{bmatrix}\begin{bmatrix} E_1 \\ 0 \end{bmatrix}. \tag{S9}$$

According to equations (S5) and (S7), equation (S9) results in $E_7 = E_9$, resulting in the indistinguishability in eavesdrwopping for both visibility and interference (see Fig. 4). Thus, the complete randomness is obtained from the MZI path superposition.

Table S1. Key distribution sequence for OKD in Fig. 1. The phase $\varphi_1$ is denoted without addition of $\varphi_2$. The mark 'X' indicates a discarded bit. The red 'X' indicates error corrections. Privacy amplification is not shown.

| Party | Sequence | Order | 1 | 2 | 3 | 4 | 5 | 6 | 7 | 8 | 9 | 10 | set |
|---|---|---|---|---|---|---|---|---|---|---|---|---|---|
| Bob | 1 | $\varphi_1$ | 0 | 0 | π | 0 | π | π | 0 | π | 0 | 0 | |
| | 2 | Prepared key: x($\varphi_1$) | 0 | 0 | 1 | 0 | 1 | 1 | 0 | 1 | 0 | 0 | {x} |
| | 8 | $V_B$ | 1 | −1 | 0.9 | 1 | −1 | −1 | −1 | 1 | 1 | −1 | |
| | 9 | raw key | X | 0 | X | X | 1 | 1 | 0 | X | X | 0 | {$m_B$} |
| | 10 | Final key | X | 0 | X | X | 1 | X | 0 | X | X | 0 | {m} |
| Alice | 3 | $V_A$ | 1 | 1 | −1 | 1 | −1 | −0.8 | 1 | −1 | 1 | −1 | |
| | 4 | Copy x: y | 0 | 0 | 1 | 0 | 1 | −0.8 | 0 | 1 | 0 | 0 | {y} |
| | 5 | $\psi_1$ | π | 0 | 0 | π | π | π | 0 | 0 | π | 0 | |
| | 6 | z($\psi_1$) | 1 | 0 | 0 | 1 | 1 | 1 | 0 | 0 | 1 | 0 | {z} |
| | 7 | raw key | X | 0 | X | X | 1 | X | 0 | X | X | 0 | {$m_A$} |
| | 10 | Final key | X | 0 | X | X | 1 | X | 0 | X | X | 0 | {m} |

Table S1 shows the regular key distribution sequence in OKD. Bob simply prepares a random bit sequence for Alice by using his basis $\varphi_1$. Alice also randomly encodes the Bob-prepared bits using her phase basis $\psi_1$. Owing to the determinacy and identity relations in MZI, the raw key is deterministically obtained by simply reading the visibilities and comparing them with their phase selections without sifting. For error corrections, both parties publically announce their error bits (red numbers in Table 1) and remove them from the raw key. As a result, the same length final key (m) is shared with each other. To evaluate an error rate, Alice and Bob randomly pick some samples out of the final key and compare them: privacy amplification. Thus, the bit rate in OKD is strongly dependent upon the bit error rate as usual. The following is an example of the key distribution sequence for OKD (see Table. S1):

1. Bob randomly selects his phase basis $\varphi_1 \in \{0, \pi\}$ to prepare a key and sends it to Alice.
2. Bob converts the chosen basis $\varphi_1$ into a key record: x∈ {0,1}; if $\varphi_1 = 0$, x=0; if $\varphi_1 = \pi$, x=1.
3. Alice measures her visibility $V_A$ and keeps the record.
4. Alice copies the Bob's key for her record y: if $V_A$=1, y=0; if $V_A$=−1, y=1; if $V_A \neq \pm 1$, y=$V_A$ (error).
5. Alice randomly selects her phase basis $\psi_1 \in \{0, \pi\}$ and sends it back to Bob.
6. Alice converts the chosen basis $\psi_1$ into a key record: z ∈ {0,1}; if $\psi_1 = 0$, z=0; if $\psi_1 = \pi$, z=1.
7. Alice compares y and z for the raw key $m_A$: $m_A = (y + z)$ at modulus 2. If $m_A \neq 0$, discard it, otherwise $m_A = z$.
8. Bob measure his visibility $V_B$.
9. Bob converts $V_B$ into his raw key m. If $V_B = -1$, $m_B = x$, otherwise discard it.
10. Alice and Bob publically announce their error bits and discard them from the raw key to set a final key.

Figure S1 shows potential network configurations of OKD, where the control phase $\varphi_2$ in Fig. 3 is assigned to the wavelength λ to support the DWDM networks.

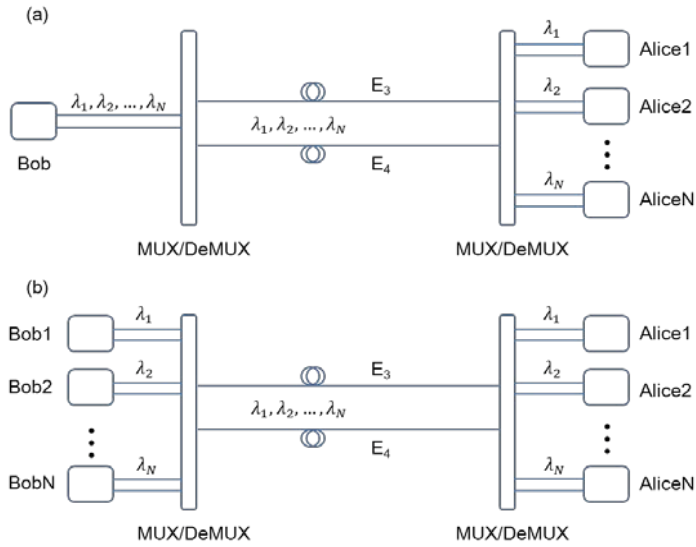

Fig. S1. A schematic of 1xN and NxN configuration for cv PKD. The $\lambda_i$ corresponds to $\varphi_2^i$ and $\psi_2^i$.